\documentclass[twocolumn,prd,aps,amsmath,amssymb,nofootinbib]{revtex4}
\usepackage{graphicx}
\usepackage{dcolumn}
\usepackage{bm}
\usepackage{mathrsfs}
\usepackage{slashed}
\usepackage{amsmath,lscape,epsfig}
\usepackage{amsfonts}
\usepackage{amsmath}
\usepackage{amssymb}
\def\beq{\begin{equation}}
\def\eeq{\end{equation}}
\def\bea{\begin{eqnarray}}
\def\eea{\end{eqnarray}}
\def\nn{\nonumber}

\def\r{\mathbf{r}}
\def\x{\mathbf{x}}

\def\z{\mathbf{z}}
\def\k{\mathbf{k}}

\begin{document}

\title{Scalar Quantum Field Theory in Disordered Media}

\author{
E. Arias$^1$\footnote{enrike@cbpf.br},
E. Goulart$^1$\footnote{egoulart@cbpf.br},
G. Krein$^2$\footnote{gkrein@ift.unesp.br},
G. Menezes$^2$\footnote{gsm@ift.unesp.br}
and N.F. Svaiter$^1$\footnote{nfuxsvai@cbpf.br}}
\affiliation{$^1$Centro Brasileiro de Pesquisas F\'{\i}sicas, Rua Dr. Xavier
Sigaud 150, 22290-180  Rio de Janeiro, RJ, Brazil\\
$^2$Instituto de F\'\i sica Te\'orica, Universidade Estadual Paulista\\
Rua Dr. Bento Teobaldo Ferraz 271 - Bloco II, 01140-070  S\~ao Paulo, SP , Brazil}

\begin{abstract}
A free massive scalar field in inhomogeneous random media is
investigated. The coefficients of the Klein-Gordon equation are
taken to be random functions of the spatial coordinates. The case
of an annealed-like disordered medium, modeled by centered stationary
and Gaussian processes, is analyzed. After performing the averages
over the random functions, we obtain the two-point causal Green's
function of the model up to one-loop. The disordered scalar
quantum field theory becomes qualitatively similar to a
$\lambda\varphi^{4}$ self-interacting theory with a
frequency-dependent coupling.
\end{abstract}


\maketitle

\section{Introduction}

Based on results obtained by Ford and collaborators~\cite{ford11,fn}
and Hu and Shiokawa~\cite{hu1}, recently three of us~\cite{krein1}
proposed an analog model for fluctuations of light cones
induced by quantum gravity effects. The model is based on two general
features of waves propagating in random fluids. First, acoustic
perturbations in a fluid define discontinuity surfaces that provide
a causal structure with sound cones. Second, propagation of
acoustic excitations in a random medium are generally described by
wave equations with a random speed of sound~\cite{ishimaru,book3,book2,book}.
Taken together, these features lead to fluctuations of sound cones,
analogous to the fluctuations of light cones. Phonons propagating in
such a random fluid are then modeling photon propagation in a gravitational
field with a fluctuating metric.

In the present paper we depart from the quest for analog models of
quantum gravity effects and study quantum fields in a disordered
environment in a more general context. Our study belongs to a wide
program devoted to propagation of quantum matter fields in a
classical background spacetime ~\cite{birrell} but with metric
fluctuations. Specifically, we consider a scalar quantum field
described by a Klein-Gordon-like equation, in which the parameters
of the equation, the mass and the coefficient of the second-order
time derivative, become random functions of the spatial coordinates.
The randomness of the parameters is due to {\em static} noise
sources that couple to the scalar field. In the limit of a vanishing
mass, one recovers the random wave equation considered in
Ref.~\cite{krein1}. While not addressing the issue of the origin of
the noise sources, we mention that one has in mind that they can be
induced by a variety of phenomena, like metric fluctuations due to
quantum creation of gravitons in a squeezed coherent state in the
presence of a black hole or interactions with background topological
defects, among others. Like in the case of quantum fields in the
presence of an external heat bath, the noise sources break Lorentz
symmetry since they define a preferred reference frame.

A scalar quantum field associated with acoustic waves in a
disordered medium can define a situation where sound cones fluctuate
randomly. In the study of such a
situation, it is important to observe that systems with disorder can
be divided into two wide groups, namely, systems with quenched or annealed
disorder ~\cite{debashish,glass,parisi1}. While in annealed systems
the random field is in thermal equilibrium with the others degrees
of freedom of the system, in quenched systems they are not.
The differentiation between the two types of disorder
is an important issue in the studies of the influence of impurities
on phase transition phenomena. Many authors have used field theory
and the renormalization group to investigate systems with quenched
disorders~\cite{lub}, with particular interest in the role played by
the disorder on critical exponents. In the case of quenched random
fields, by means of the replica trick it is possible to define a
quenched generating functional of connected $n$-point functions.
With this generating functional in hand, the issue of the
influence of the disorder on phase transitions can be studied. In
the present paper we consider annealed disorder. We consider weak
noise fields and implement a perturbative expansion controlled by
small parameters that characterize the noise correlation functions.
We obtain causal two- and four-point Green's functions of the scalar
field. Averaging these Green's functions over the noise
fluctuations, one obtains two- and four-point functions
qualitatively similar to a self-interacting $\lambda\varphi^{4}$
theory. As will be shown, we obtain a frequency-dependent coupling
constant.

Since Unruh's original paper~\cite{unruh1}, the possibility of
simulating aspects of general relativity and quantum fields in
curved space-time through analog models has been widely discussed in
the literature~\cite{mario,lectures}. One interesting proposal is
the generation of an acoustic metric in Bose-Einstein condensates
and superfluids~\cite{am7}. On the other hand, analog models with
sonic black-hole could find a sort of generalization within the
random fluid scenario. Since Bose-Einstein condensates are natural
candidates to produce an acoustic black-hole, one way to go beyond
the semi-classical approximation is to inquire how these systems
behave in the presence of disorder. The calculations presented in
the present paper are the first steps in the implementation of such
a program.

The organization of the paper is as follows: In Section~\ref{sec:PT} we
discuss the perturbative approach in a free scalar field in the
presence of annealed disorder described by two random functions.
Section III contains our conclusions. In the appendix some lengthy
calculations are presented.

\section{Perturbation Theory in Annealed-like Disordered Media}
\label{sec:PT}

\vspace{0.5cm}

Let us consider a scalar field $\varphi(t, \r)$, defined in a $(d+1)$ dimensional
space-time, that
satisfies the random Klein-Gordon equation:
\begin{eqnarray}
\left[\left(1 + \mu \right)\frac{1}{u_{0}^{2}}
\frac{\partial^{2}}{\partial\,t^{2}}-\nabla^{2}+
\left(1 + \xi \right) m^{2}_{0} \right]
\varphi(t,{\r})=0.
\label{KG-eq}
\end{eqnarray}
The quantities $\mu = \mu(\r)$ and $\xi = \xi(\r)$ are dimensionless random variables,
functions of the spatial coordinates only, whose statistical properties will
be defined shortly. Here, the random coefficient of the time derivative
characterizes the possible discontinuity surfaces of the theory and provides a
kind of random causal structure to it. It is well known that the
principal part of a differential equation, i.e. the terms with
the higher-order derivatives, determines entirely the \textit{loci}
of points in space-time where a solution may possess non-null
discontinuities. This property is at the root of all analogue models
of classical gravitation, namely, an effective causality can be
obtained from the kinematical properties of a physical system.
The region of influence of excitations is given by an envelope which
characterizes the maximum speed of propagation, that is, the light cone
of the theory. From a physical point of view the region of influence of the
theory may be obtained by an \textit{eikonal} approximation, where
the mass term may be disregarded. On the other hand, when the wave
frequency is of the same magnitude of the mass term, excitations of
the system propagate inside the characteristic cone.

In the present paper we consider zero-mean random functions
$\mu(\r)$ and $\xi(\r)$:
\begin{equation}
\langle \mu(\r) \rangle_{\mu} = 0, \hspace{0.5cm}
\langle \xi(\r) \rangle_{\xi} = 0,
\label{means}
\end{equation}
and, for simplicity, we suppose white-noise correlations:
\begin{eqnarray}
\langle \mu(\r) \mu(\r') \rangle_{\mu} &=& \sigma^2_\mu \,
\delta(\r - \r') ,
\label{corr-mu}
\\
\langle \xi(\r) \xi(\r') \rangle_{\xi} &=& \sigma^2_\xi \,
\delta(\r - \r') ,
\label{corr-xi}
\end{eqnarray}
where $\langle\, \cdots \,\rangle_{\mu}$ and $\langle\, \cdots \,\rangle_{\xi}$
denote ensemble average of noise realizations and $\sigma_\mu$ and
$\sigma_\xi$ characterize the strengths of the noises. We also suppose
that the noises are Gaussian distributed (we use the notation $\mu(\r_i)
= \mu_i $):
\begin{eqnarray}
\langle\mu_{i_1}  \cdots \mu_{i_{2n}} \rangle_\mu  &=&
\langle \mu_{i_1} \mu_{i_2} \rangle_\mu \langle \mu_{i_3} \mu_{i_4} \rangle_\mu
  \cdots \langle \mu_{i_{2n-1}} \mu_{i_{2n}} \rangle_\mu \nonumber \\
&& + \, {\rm permutations} ,
\label{wick}
\end{eqnarray}
and correlations of an odd number of noises are zero. The reader may
be aware, but it is worth remembering that these random functions
$\mu$ and $\xi$ are statistically independent.

\begin{figure}[htb]
\centering
\includegraphics[scale=0.17]{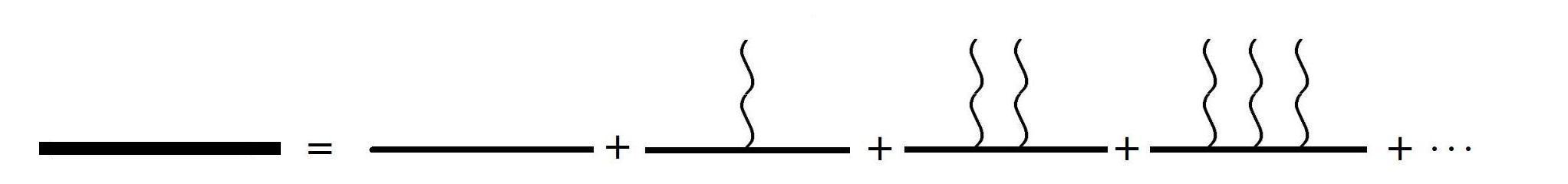}
\caption{Perturbative expansion of $G$ in terms of the
disorder. The wavy lines represent generically the random functions
$\mu$ and $\xi$.}
\label{expansion}
\end{figure}

Since Eq.~(\ref{KG-eq}) is linear in $\varphi(t,\r)$, it is
useful to resort to Fourier transforms in order to find its
solutions. Therefore, we define the Fourier transform of
$\varphi(t,\r)$ on the time variable $t$:
\begin{equation}
\varphi(t,\r\,) = \int \frac{d\omega}{2\pi} \, e^{-i \omega t}\,
\varphi(\omega,\r), \label{FTt-phi}
\end{equation}
and on the space variable $\r$:
\begin{equation}
\varphi(\cdot ,\r ) = \int \frac{d\k}{(2\pi)^d} \,
e^{i \k \cdot \r } \, \varphi(\cdot,\k).
\label{FTr-phi}
\end{equation}
In addition, we define Fourier transforms of the stationary noise
functions:
\begin{equation}
\mu(\r ) = \int \frac{d\k}{(2\pi)^d} \,
e^{i \k\cdot\r } \,
\mu(\k ) ,
\label{FT-noise}
\end{equation}
and similarly for $\xi(\r)$. Using these in Eq.~(\ref{KG-eq}),
one obtains that the Fourier components of the field
$\varphi(\omega,\k)$ satisfy an algebraic equation that can
be written as
\begin{equation}
\int d\k' \, \left[L_0(\k,\k') + L_1(\k,\k')\right] \, \varphi(\omega,\k') = 0,
\label{KG-mom}
\end{equation}
where $L_0$ is a non-stochastic matrix with elements:
\begin{eqnarray}
L_{0}(\k,\k') = \left(\frac{\omega^{2}}{u_{0}^{2}}
- \k^{2} - m^{2}_{0}\right) \delta(\k -\k'),
\label{L0-mom}
\end{eqnarray}
and $L_1(\k,\k')$ is a stochastic matrix, with elements:
\begin{eqnarray}
L_{1}(\k,\k') = \frac{1}{(2\pi)^d}\left(  \mu(\k -\k') \, \frac{\omega^2}{u^2_0} -
\xi(\k -\k') \, m^{2}_{0}\right).
\label{L1-mom}
\end{eqnarray}
In $\r$-space, $L_0$ and $L_1$ read:
\begin{eqnarray}
L_{0}(\r) &=& \frac{\omega^{2}}{u_{0}^{2}} + \nabla^{2} - m^{2}_{0},
\label{nami32} \\
L_{1}(\r) &=& \mu(\r)
\frac{\omega^{2}}{u_{0}^{2}}  - \xi(\r)\,m^{2}_{0} .
\label{nami33}
\end{eqnarray}
As in $\k$-space, they act as integral convolution operators. Note
that while $L_{1}$ is diagonal, $L_{0}$ is non-diagonal in
$\r$-space; yet, in $\k$-space the situation is the opposite. In
terms of these operators, the random Klein-Gordon equation can be
written in matrix form:
\begin{equation}
(L_{0}+\,L_{1}) \, \varphi(\omega,\cdot) = 0 .
\label{nami31}
\end{equation}
From this, one can define the full (operator valued) Green's function
$G$:
\begin{equation}
G=( L_{0} + L_{1} )^{-1}.
\label{green}
\end{equation}
Now, if one assumes that the noises are ``weak",
a natural perturbative
expansion for~$G$ in the form of a Dyson formula can be defined:
\begin{eqnarray}
G &=& G^{(0)} - G^{(0)}\, L_1 \,G^{(0)} + G^{(0)}\, L_1 \,G^{(0)}\, L_1 \,G^{(0)}
+ \, \cdots \nonumber \\
&=& G^{(0)} - G^{(0)} \, \Sigma \, G^{(0)} ,
\label{green1}
\end{eqnarray}
with the self-energy $\Sigma$ given by
\begin{equation}
\Sigma = L_1 - L_1 G_0 L_1 + \cdots \, ,
\label{self}
\end{equation}
where $G^{(0)}=\,L_{0}^{-1}$ is the unperturbed (operator valued) Green's
function which, in $\k$-space, can be written as
\begin{equation}
G^{(0)}(\omega,\k) = \frac{1}{\omega^2 - (\k^2 + m^2_0) + i\epsilon} .
\label{nami34}
\end{equation}
(Hereafter we take $u_0$ equal to unity.) A schematic representation
of the expansion in Eq.~(\ref{green1}) is shown in
Fig.~\ref{expansion}.

We note that $i G^{(0)} = \Delta_0$, where $\Delta_0$ is the noninteracting Feynman
propagator, which is the vacuum expectation value of the time ordered product
of the quantum field operators $\varphi$ at space-time points
$x=(t,\x)$ and $x'=(t',\x')$:
\beq
\Delta_0(x,x') = \langle 0|T[\varphi(t,\x) \varphi(t',\x')]|0\rangle .
\label{defDelta}
\eeq

Let us write down explicitly a few terms of the perturbative
series in $\r$-space, up to second order in the random fields:
\begin{widetext}
\begin{eqnarray}
G(\omega,\r,\r') &=& G^{(0)}(\omega,\r,\r') -
\int d\r_1 \, G^{(0)}(\omega,\r,\r_1)
\left[\mu(\r_1) \,\omega^2 - \xi(\r_1)\,m^2_0 \right]
G^{(0)}(\omega,\r_1,\r')  \nonumber \\
&+&  \int d\r_1 \int d\r_2 \, G^{(0)}(\omega,\r,\r_2)
\left[\mu(\r_{2}) \, \omega^{2} - \xi(\r_{2})\,m^{2}_{0} \right]
\, G^{(0)}(\omega,\r_{2},\r_{1}) \nonumber\\
&\times& \left[\mu(\r_{1}) \, \omega^{2} - \xi(\r_{1}) \, m^{2}_{0} \right]
\, G^{(0)}(\omega,\r_{1},\r') + \cdots \,.
\label{expans}
\end{eqnarray}
\end{widetext}
A pictorial representation in $\r$-space of this perturbative expression
in terms of diagrams can be done. They correspond to diagrams of multiple-scattering of the
Klein-Gordon field on random inhomogeneity scatterers located at positions
$\r_{1}, \r_{2},\,\cdots \,$. In $\k$-space, similar diagrams correspond to
multiple interactions of Fourier components of the Klein-Gordon field and
of the random inhomogeneities.

\begin{figure}[h]
\centering
\includegraphics[scale=0.3]{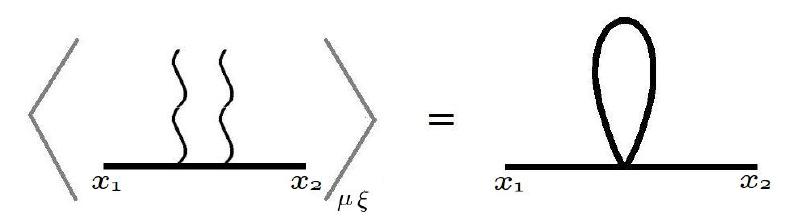}
\caption{The one-loop correction to the two-point causal function
due to the random inhomogeneities.}
\label{selfenergy}
\end{figure}

Let us now perform the averages over the random processes that appear in the
definition of the propagator $G$ given in Eq.~(\ref{expans}). Note that, because of
the Gaussian nature of noise correlations, terms with an odd-number
of noise fields do not contribute to the two-point function, and the
first nonzero correction to the two-point function comes from the
averages over the third term in Eq.~(\ref{expans}). Therefore,
up to second order in the noise fields, we get
\begin{eqnarray}
{G}^{(1)}(\omega,\r,\r') & \equiv & \langle \, G(\omega,\r,\r') \rangle_{\mu\xi}
\nonumber \\
&=& G^{(0)}(\omega,\r,\r') + \bar G^{(1)}(\omega,\r,\r'),
\label{onelooppropagator}
\end{eqnarray}
with the one-loop correction $\bar G^{(1)}(\omega,\r,\r')$, represented pictorially
in Fig.~\ref{selfenergy} and given in terms of its Fourier transform
$\bar G^{(1)}(\omega,\k)$ as
\begin{equation}
\bar G^{(1)}(\omega,\r,\r') = \int \frac{d\k}{(2\pi)^d} \,
\bar G^{(1)}(\omega,\k) \, e^{i\k\cdot (\r -\r')}.
\end{equation}
This quantity can be written in terms of a self-energy $\Sigma(\omega,\k)$ --
defined consistently with Eq.~(\ref{green1}) -- as:
\begin{equation}
\bar G^{(1)}(\omega,\k) = - G^{(0)}(\omega,\k) \, \Sigma(\omega,\k) \,
G^{(0)}(\omega,\k) ,
\end{equation}
with
\begin{equation}
\Sigma(\omega, \k) = - \left(\sigma^2_\mu \omega^4
+ \sigma^2_\xi  m_0^4\right) \, \alpha(\omega) ,
\label{Sigma}
\end{equation}
where
\begin{equation}
\alpha(\omega) = \int \frac{d\k}{(2\pi)^d} \, G^{(0)}(\omega,\k) .
\label{alpha}
\end{equation}
Note that at one-loop order, the self-energy is actually independent
of $\k$. From Eq.~(\ref{green1}), one has $G^{-1} = G^{(0)-1} +
\Sigma$ and, therefore, in the $\k$-representation
\begin{eqnarray}
[G^{(1)}(\omega,\k)]^{-1} &=& [G^{(0)}(\omega,\k)]^{-1} + \Sigma(\omega,\k)
\nonumber \\
&=& \omega^ 2 - \k^2 - m^2,
\end{eqnarray}
where
\begin{eqnarray}
m^2 &=& m^2_0 - \Sigma(\omega,\k) \nonumber \\
&=& m^2_0 + \left(\sigma^2_\mu \omega^4
+ \sigma^2_\xi  m_0^4\right) \, \alpha(\omega) .
\label{mass}
\end{eqnarray}

\vspace{0.25cm}
From this result, it is clear that the effect of randomness is to
turn a free, conventional scalar quantum field theory into a
self-interacting theory, with a self-interaction qualitatively
similar to $\lambda \varphi^{4}$. The induced $\lambda \varphi^{4}$
theory is a consequence of the Gaussian nature of the functions
$\mu(\vec{r}\,)$ and $\xi(\vec{r}\,)$; a more general polynomial
self-interacting model is obtained if non-Gaussian noises are used.
The integral is divergent because of the white-noise nature of the
noise; a colored noise correlation function would lead to a finite
integral.

As mentioned earlier in the text, the induced coupling generated is
frequency-dependent, $\lambda(\omega) \sim \sigma^2_\mu \omega^4 +
\sigma^2_\xi  m_0^4$. Also, for $d=3$ one can isolate the finite
part from the integral making use of the identity
\beq \frac{|\k|^2}{|{\k}|^2 - (\omega^2 - m^2_0)} = 1 +
\,\frac{\omega^2 - m^2_0}{|\k|^2 - (\omega^2 - m^2_0)},
\label{separat} \eeq
so that the finite part of the self-energy, $\Sigma_f(\omega,\k)$,
is given by
\bea \Sigma_{f}(\omega,\k) &=&  \left(\sigma^2_\mu \omega^4 +
\sigma^2_\xi  m_0^4\right) \int^{\infty}_0 \frac{d|\k|}{2\pi^2}\,
\frac{\omega^2 - m^2_0}{\omega^2 - |{\k}|^2 - m^2_0} \nonumber \\
&=&  \frac{1}{4\pi}\left(\sigma^2_\mu \omega^4
+ \sigma^2_\xi  m_0^4\right) |\omega^2 - m^2_0|^{1/2} {\cal A},
\eea
with
\beq
{\cal A} = \Biggl\{
\begin{array}{ll}
+1\,, & \hspace{0.25cm} m^2_0  >  \omega^2 ,\\
-i \,,
& \hspace{0.25cm} \omega^2  > m^2_0 .
\end{array}
\label{f}
\eeq

The conclusion from this calculation is that the random fluctuations
induce a self-energy with a width, which gives a finite life-time
for the excitations.

Since randomness, as mentioned previously, led to an induced
coupling, one may ask on the form of the perturbative corrections to
the four-point function $G_4$. The reason for this particular
interest is the following. As well known, given the four-point
function, one can define the one-particle-irreducible ($1$PI)
four-point connected Green's function without the external legs
$\Gamma_4$:
\begin{equation}
\Gamma_{4}(k_1,k_2,k_3,k_4) =
G_{4}(k_1,k_2,k_3,k_4)\prod^4_{i=1}\Gamma_{k_i}, \label{vertex}
\end{equation}
where
\beq \Gamma_k = \omega^2-(\k^2 + m_0^2). \label{Gamma-k} \eeq
In a covariant quantum field theory, one may compute the
renormalized coupling constant from the proper vertex function,
under certain conditions. So, even though we are not dealing with a
covariant field theory, it is natural to expect that such function
can bring us some information on the nature of the induced coupling.

\begin{figure}[h]
\includegraphics[scale=0.19]{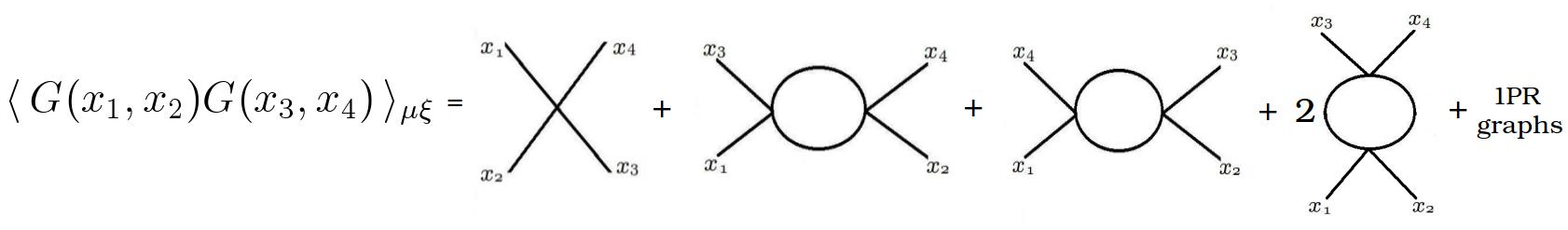}
\caption{The four-point Green function up to one-loop, after noise
averaging. `1PR graphs' means one-particle reducible
graphs.}
\label{g12g34}
\end{figure}

Since the random field equation is linear in the field variable, the
corresponding action of the system is quadratic in the field,
resulting in a Gaussian generating functional. Then, the quantum
$n$-point Green's functions of the system are products of two-point
Green's functions. In particular the four-point Green's function
will be of the form:
\begin{eqnarray}
G_4(x_1,x_2,x_3,x_4) &=& \langle G(x_1,x_2)G(x_3,x_4)
\rangle_{\mu\xi} \nonumber\\
&+& \langle G(x_1,x_3)G(x_2,x_4) \rangle_{\mu\xi} \nonumber\\
&+& \langle G(x_1,x_4)G(x_2,x_3) \rangle_{\mu\xi}. \label{g4a}
\end{eqnarray}
As before, $x = (t,\x)$. Fig.~\ref{g12g34} presents a graphical
representation of this expression.

\begin{figure}[h]
\centering
\includegraphics[scale=0.25]{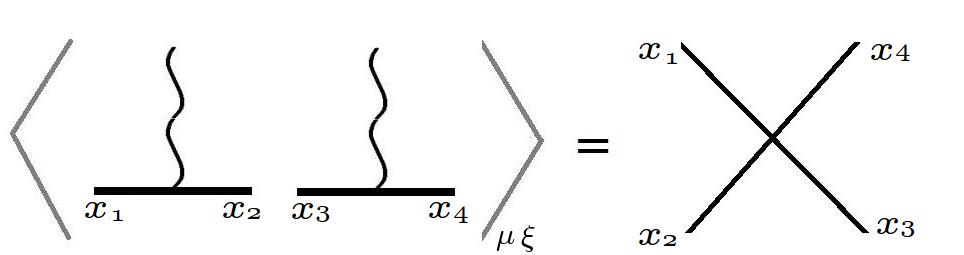}
\caption{The disorder-induced, tree-level four-point Green's
function.}
\label{treelevel4point1}
\end{figure}

\begin{figure}[h]
 \centering
 \includegraphics[scale=0.25]{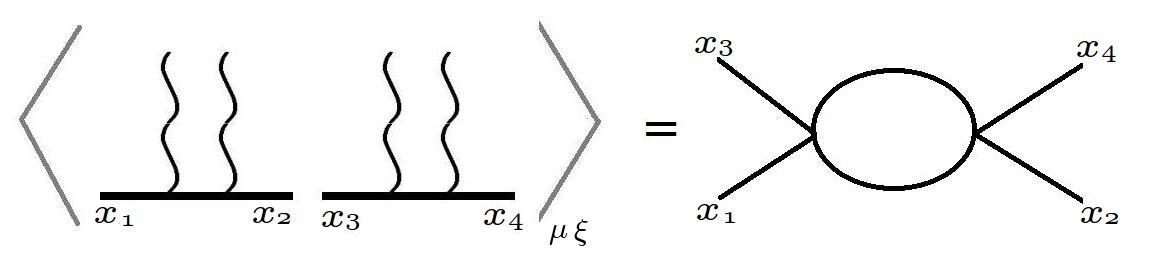}
\caption{One-loop corrections of first kind to the four-point Green's function.}
\label{oneloop4point}
\end{figure}

\begin{figure}[h]
\includegraphics[scale=0.29]{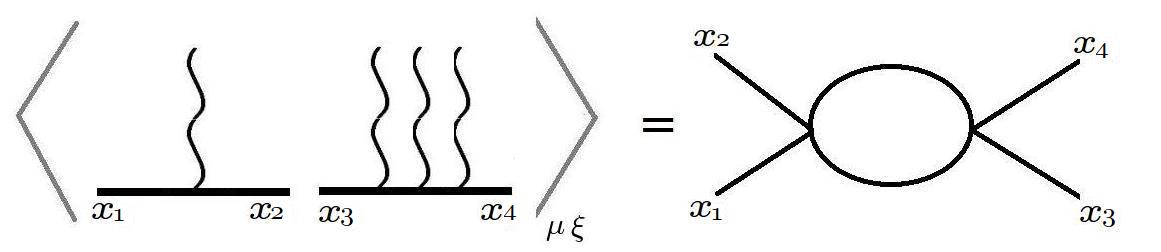}
\caption{One-loop corrections of second kind to the four-point Green's function.}
\label{gra_apend}
\end{figure}

In Appendix~\ref{app:G2-G4}, Eq.~(\ref{g4a}) is computed up to one
loop; the result can be written as $G_4(x_1,x_2,x_3,x_4) =
G_4^{(0)}(x_1,x_2,x_3,x_4)  +   G_4^{(1)}(x_1,x_2,x_3,x_4)$, where
$G_{4}^{(0)}$ and $G_4^{(1)}$ are the tree-level and one-loop
contributions, respectively. The noise averaging leading to the
tree-level four-point function $G_{4}^{(0)}$ is illustrated in
Fig.~\ref{treelevel4point1} and the one-loop
corrections $G_4^{(1)}$ are pictorially represented in Figs.~\ref{oneloop4point} and \ref{gra_apend}.

From Eq. (\ref{vertex}) and the results derived in the Appendix, one
can express the tree-level proper vertex function as:
\bea \Gamma^{(0)}_4(k_1,k_2,k_3,k_4) &=&
 (2\pi)  \Bigl[ \delta(\omega_3 + \omega_4)
\left(\sigma^2_\mu \omega^2_2 \omega^2_4 +
\sigma^2_\xi m^4_0\right) \nn \\
&+& \delta(\omega_2 + \omega_4) \left(\sigma^2_\mu \omega^2_3
\omega^2_4 + \sigma^2_\xi m^4_0\right)\nn \\ &+& \delta(\omega_2 +
\omega_3) \left(\sigma^2_\mu \omega^2_3 \omega^2_4 + \sigma^2_\xi
m^4_0\right) \Bigr]. \label{Gamma4-0}
\end{eqnarray}

As mentioned earlier, the proper understanding of the nature of the induced interacting
theory requires the study of the four-point Green's function.
We will not dwell into such a study here. Besides being involved (see the
Appendix for the one-loop contribution), such a study is out of the
scope of the present paper and therefore we reserve it for a future
publication.

The most striking feature that comes out from these calculations is
that the induce coupling is qualitatively similar to a $\lambda
\phi^4$ theory, with a frequency-dependent coupling. The important
meaning of this is that if one starts with an interacting model, the
induced coupling due to the random scatterings in the medium changes
the value of the renormalized coupling constant of the model. As a
result, a relevant question to be answered is to know what would be
the sign of this effective coupling. For instance, if it turns out
to be negative, it is clear that we would have a decrease in the
value of the renormalized coupling constant. Furthermore, as well
known, a negative coupling constant can be a source of instability
in field theory, which means that one should look for alternative
solutions to circumvent the above mentioned problem. One possibility
is to discuss the ground state properties of the system in a theory
with metastable vacua. This can be obtained in non-simply connected
manifolds. For simplicity one might assume periodic boundary
conditions in all $d+1$ directions, with compactified lengths $L_1$,
$L_2$,..,$L_{d+1}$. At this point we remark that in theories defined
on a non-simply connected Euclidean space, one of the compactified
length is related to the temperature, remembering the
Kubo-Martin-Schwinger (KMS) condition \cite{matsu}. Therefore in a
Euclidean theory describing bosons, we have to impose periodic
boundary condition in one compactified direction. In the other $d$
directions, we are free to choose any boundary conditions. However,
imposing periodic boundary conditions in all compactified dimensions
enables us to maintain translational invariance~\cite{ford}.
Non-translational invariant systems were studied in Refs. \cite{sy}.
We take this opportunity to call attention for the results obtained
by Arias and co-workers \cite{arias}, where the thermodynamic of the
massless self-interacting scalar field model with negative coupling
constant was analyzed. We believe that such results from this
reference can be useful to proceed with the investigations of the
consequences of impurities in a relativistic model.

\section{Conclusions}
\label{sec:concl}

Recently it was proposed that an analog model for quantum
gravity effects can be implemented in condensed matter physics when
the fluid is a random medium \cite{krein1}. In order to study the
wave propagation of the elementary excitations in the fluid when the
acoustic wave propagates in random medium, it was considered that
there are no random fluctuation in the density, but only in the
reciprocal of the bulk modulus. In this analog model of quantum
gravity with a colored noise a free quantum field theory describing
phonons becomes a self-interacting model.

In this paper we studied a massive scalar field theory in a
disordered medium. We determined the perturbation theory in
annealed-like disordered medium, with two distinct random functions.
After performing the random averages over the noise function a free
scalar quantum field theory became the $\lambda\varphi^{4}$
self-interacting model.
The one-loop two and four-point functions were presented. We obtained
that the induced coupling constant is frequency dependent. Note
that although we employed Gaussian
functions, it is not difficult to extend the method to non-Gaussian
functions by means of a cluster expansion \cite{cluster}.

There exist several possibilities to use the above results in analog
models. For example, a random fluid with a supersonic acoustic flow
can be an analog model that allows one to discuss the effect of the
fluctuation of the geometry in the Hawking radiation. Since
Bose-Einstein condensate is a natural candidate to produce acoustic
black-holes, in order to introduce randomness, i.e., the effects of
the metric fluctuations in the acoustic black-hole, we have to
investigate how superfluid and Bose-Einstein condensate behave in
the presence of disorder. Sound waves in Bose-Einstein condensate,
described by random wave equation must reproduce physics beyond the
semi-classical approximation in an analog model.

An interesting course of action would be the search of a possible
localization of the Hawking thermal flux in such analog model, since
it is well known that the effect of the impurities is the
localization of classical waves and elementary excitations
\cite{anderson}. In conclusion, the study of quantum fields in
disorder medium in the analog models scenario introduces new
experimental and theoretical challenges. The study of the
relativistic Bose-Einstein condensation in the presence of a random
potential \cite{new}, and also the consequences of introducing
metric fluctuations in the analog model proposed in Ref.~\cite{Fagnocchi:2010sn}
is under investigation by the authors.


\acknowledgments
Work partially financed by CAPES, CNPq and FAPESP (Brazilian agencies).

\appendix

\section{Detailed derivation of the causal two- and four-point functions}
\label{app:G2-G4}

\begin{figure}[h]
\centering
\includegraphics[scale=0.4]{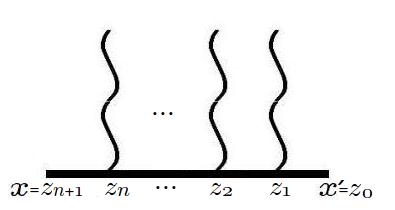}
\caption{Graphical representation of a generic term in Eq.~(\ref{genericterm})}
\label{generic}
\end{figure}

We start from the perturbative expansion for the two-point function ~$G$,
Eq.~(\ref{green1}), which we rewrite as:
\begin{eqnarray}
G &=& G^{(0)} - G^{(0)}\, L_1 \,G^{(0)} + G^{(0)}\, L_1 \,G^{(0)}\, L_1 \,G^{(0)}
+ \, \cdots \nonumber \\
&=& G^{(0)} \left(1 + \sum^\infty_{n=1} {\cal G}^{(n)} \right),
\label{green1-app}
\end{eqnarray}
with ${\cal G}^{(n)}$ given by
\beq
{\cal G}^{(n)} = (-1)^{n} \prod_{j=1}^n \left(L_1 \, G^{(0)}\right)^j ,
\label{calG}
\end{equation}
In terms of time and space variables, $x=(t,\x)$, this corresponds
to
\begin{equation}
G(x,x') = \int dz_1 \,G^{(0)}(x-z_1)\biggl[\delta(z_1-x') +
\sum_{n=1}^{\infty}{\cal G}^{(n)}(z_1,x')\biggr],
\label{a1}
\end{equation}
and
\begin{equation}
{\cal G}^{(n)}(z_1,x') = (-1)^{n}\prod_{j=1}^n L_1(z_j)
\int dz_{j+1} \,G^{(0)}(z_j,z_{j+1}),
\label{genericterm}
\end{equation}
where $L_1 (x)$ is given by
\beq
L_1 (x) = L_1(t,\x) = - \mu(\x) \frac{\partial^2}{\partial t^2}
- \xi(\x) m_0^2 .
\label{L1-coord}
\eeq
In terms of the frequency and wave-number coordinates in Fourier
space, $(\omega,\k)$, $L_1$ is given by Eq.~(\ref{L1-mom}). In
Eq.~(\ref{genericterm}), it is to be understood that $z_{n+1} = x'$
and that there is no integration in $z_{n+1}$. In Fig.~\ref{generic}
we present a graphical representation of a generic term in equation
(\ref{genericterm}).

For completeness, we give here the expression for Fourier transform of the free
Green's function $G^{(0)}(x-x')$:
\beq
G^{(0)}(x,x') = \int \frac{dk}{(2\pi)^{d+1}} \,e^{-ik(x-x')} \, G^{(0)}(k) ,
\eeq
with
\bea
G^{(0)}(k) = G^{(0)}(\omega,\k) =  \Gamma_k^{-1} =
\frac{1}{\omega^2-(\k^2 + m_0^2)}.
\eea
From the properties of the noise functions, given in
Eqs.~(\ref{means})-(\ref{corr-xi}), one has that:
\begin{equation}
\langle L_1(x)\rangle_{\mu\xi}=0,
\end{equation}
and
\begin{equation}
\langle L_1(x) L_1(x')\rangle_{\mu\xi} = \left(\sigma^2_\mu
\frac{\partial^2}{\partial t^2} \, \frac{\partial^2}{\partial t'^2}
+ \sigma^2_\xi \, m_0^4\right) \delta(\x -\x'), \label{avLL}
\end{equation}
From this, it is clear that:
\beq
\langle{\cal G}^{(2n+1)}(x,x')\rangle_{\mu\xi}=0,
\eeq
and, therefore, the first correction to the two-point causal Green's
function is given by:
\beq
\bar G^{(1)}(x,x') = \int dz_1 \, G^{(0)}(x,z_1)
\,  \langle {\cal G}^{(2)}(z_1,x')\rangle_{\mu\xi} .
\eeq
Now, for $n=2$, Eq.~(\ref{genericterm}) gives:
\bea
\hspace{-0.1cm}
\langle {\cal G}^{(2)}(z_1,x')\rangle_{\mu\xi} &=&
\int dz_{2} \, \langle L_1(z_1) L_1(z_2) \rangle_{\mu\xi}
\nn\\
&\times&  G^{(0)}(z_1,z_2) G^{(0)}(z_2,x').
%
\eea
Multiplying this by $G^{(0)}(x-z_1)$ and integrating over $z_1$,
one easily obtains the result:
\begin{equation}
\bar G^{(1)}(x,x') = \int \frac{dk}{(2\pi)^{d+1}} \,
e^{-ik(x-x')} \, \bar G^{(1)}(k),
\end{equation}
with
\bea
\bar G^{(1)}(k) &=& \bar G^{(1)}(\omega,\k) \nn\\
&=& - G^{(0)}(\omega,\k) \Sigma(\omega,\k) G^{(0)}(\omega,\k), \eea
where $\Sigma(\omega,\k)$ is the self-energy given in
Eq.~(\ref{Sigma}).

Next, we derive the four-point function. Since the full action of the model
is quadratic in the field, the
quantum four-point function can be factored into the product of
two two-point functions. The average over the noises of this function
can then be written as
\begin{eqnarray}
G_4(x_1,x_2,x_3,x_4) &=& \langle G(x_1,x_2)G(x_3,x_4) \rangle_{\mu\xi}
\nonumber\\
&+& \, \langle G(x_1,x_3)G(x_2,x_4) \rangle_{\mu\xi}
\nn \\
&+& \, \langle G(x_1,x_4)G(x_2,x_3) \rangle_{\mu\xi} .
\label{g4}
\end{eqnarray}
We are interested only in the one-particle irreducible (1PI) parts of
this. The lowest-order 1PI contribution is the tree-level four-point
function -- see Fig.~\ref{treelevel4point1}. For first term in
Eq.~(\ref{g4}) one has:
\bea
\hspace{-0.25cm}
\langle G(x_1,x_2)G(x_3,x_4) \rangle^{(0)}_{\mu\xi}
&=& \int dz_1 \int dz_2 \, G^{(0)}(x_1,z_1)
\nn \\
&&\hspace{-3.25cm}\times
\, G^{(0)}(x_3,z_2) \langle\,{\cal G}^{(1)}(z_1,x_2)
{\cal G}^{(1)}(z_2,x_4)\,\rangle_{\mu\xi}.
\eea
For $n=1$, Eq.~(\ref{genericterm}) gives:
\bea
\langle {\cal G}^{(1)}(z_1,x_2) {\cal G}^{(1)}(z_2,x_4) \rangle_{\mu \xi}
&=& \langle L_1(z_1) L_1(z_2) \rangle_{\mu\xi}  \nn \\
&&\hspace{-1.5cm}\times \, G^{(0)}(z_1,x_2) G^{(0)}(z_2,x_4)   .
\eea
Using the result of Eq.~(\ref{avLL}) and performing the appropriate Fourier
transforms, one obtains:
\begin{widetext}
\bea
\langle G(x_1,x_2)G(x_3,x_4) \rangle^{(0)}_{\mu\xi}
&=& \int \frac{dk_1}{(2\pi)^{d+1}} \cdots
\int \frac{dk_4}{(2\pi)^{d+1}} \, e^{-i( k_1 x_1 - k_2 x_2 + k_3 x_3 - k_4 x_4)}
G^{(0)}(k_1) G^{(0)}(k_2) G^{(0)}(k_3) G^{(0)}(k_4)
\nn\\
&\times& \, \left(\sigma^2_\mu \omega^2_2 \omega^2_4 +
\sigma^2_\xi m^4_0\right) \int dz_1 \int dz_2 \, e^{i (k_1 - k_2) z_1 + i (k_3 - k_4) z_2}
\, \delta(\z_1 - \z_2)  \nn \\
&=& \int \frac{dk_1}{(2\pi)^{d+1}} \cdots
\int \frac{dk_4}{(2\pi)^{d+1}} \, e^{-i( k_1 x_1 + k_2 x_2 + k_3 x_3 + k_4 x_4)}
G^{(0)}(k_1) G^{(0)}(k_2) G^{(0)}(k_3) G^{(0)}(k_4)
\nn\\
&\times& \, \left(\sigma^2_\mu \omega^2_2 \omega^2_4 + \sigma^2_\xi
m^4_0\right)(2\pi)^{d+1} \delta(k_1 + k_2 + k_3 + k_4)
(2\pi)\delta(\omega_3 + \omega_4)
 .
\label{G4-0-1}
\end{eqnarray}
To arrive at the last result, we have made the changes $k_2 \rightarrow - k_2$ and
$k_4 \rightarrow - k_4$, and used the identity $\delta(\k_1 + \k_2 + \k_3 + \k_4)
\delta(\omega_1 + \omega_2) \delta(\omega_3 + \omega_4) = \delta(k_1 + k_2 + k_3 + k_4)
\delta(\omega_3 + \omega_4)$. The other tree-level contributions in Eq.~(\ref{g4}) can be obtained from
the result in Eq.~(\ref{G4-0-1}) by appropriate changes of integration variables.
Specifically, the final result for the tree-level four-point function,
$G^{(0)}_4(x_1,x_2,x_3,x_4)$, can be written as:
\bea G^{(0)}_4(x_1,x_2,x_3,x_4) &=& \int \frac{dk_1}{(2\pi)^{d+1}}
\cdots \int \frac{dk_4}{(2\pi)^{d+1}} (2\pi)^{d+1} \delta(k_1 + k_2
+ k_3 + k_4)\, \nn \\ &\times& e^{-i( k_1 x_1 + k_2 x_2 + k_3 x_3 +
k_4 x_4)}\, G^{(0)}_4(k_1,k_2,k_3,k_4), \eea
with
\bea
G^{(0)}_4(k_1,k_2,k_3,k_4) =
G^{(0)}(k_1) G^{(0)}(k_2) G^{(0)}(k_3) G^{(0)}(k_4) \,
\Gamma^{(0)}_4(k_1,k_2,k_3,k_4) ,
\eea
where
\bea
\Gamma^{(0)}_4(k_1,k_2,k_3,k_4) &=&
 (2\pi)  \Bigl[ \delta(\omega_3 + \omega_4)
\left(\sigma^2_\mu \omega^2_2 \omega^2_4 +
\sigma^2_\xi m^4_0\right) \nn \\
&+& \delta(\omega_2 + \omega_4) \left(\sigma^2_\mu \omega^2_3
\omega^2_4 + \sigma^2_\xi m^4_0\right) + \delta(\omega_2 + \omega_3)
\left(\sigma^2_\mu \omega^2_3 \omega^2_4 + \sigma^2_\xi m^4_0\right)
\Bigr]. \label{Gamma4-0}
\end{eqnarray}

Next, let us consider the one-loop corrections to the four-point
function. Let us focus on contractions which are depicted in
Fig.~\ref{oneloop4point}. For the first term in Eq. (\ref{g4}) we have:
\bea \langle G(x_1,x_2)G(x_3,x_4) \rangle^{(1-I)}_{\mu\xi} &=& \int
dz_1 \int dz_3 \, G^{(0)}(x_1,z_1) G^{(0)}(x_3,z_3) \langle\,{\cal
G}^{(2)}(z_1,x_2) {\cal G}^{(2)}(z_3,x_4)\,\rangle_{\mu\xi}.
\label{loop}
\eea
From Eq. (\ref{genericterm}), one has:
\bea
\langle\,{\cal G}^{(2)}(z_1,x_2)
{\cal G}^{(2)}(z_3,x_4)\,\rangle_{\mu\xi} &=&
\int dz_{2} \int dz_4\, \langle L_1(z_1) L_1(z_2) L_1(z_3) L_1(z_4)
\rangle_{\mu\xi}
\nn\\
&\times&
\, G^{(0)}(z_1,z_2) G^{(0)}(z_2,x_2)  G^{(0)}(z_3,z_4) G^{(0)}(z_4,x_4).
\eea
Using now the fact that the noise correlations are Gaussian :
\bea \langle L_1(z_1) L_1(z_2) L_1(z_3) L_1(z_4) \rangle_{\mu\xi}
&=& \langle L_1(z_1) L_1(z_3) \rangle_{\mu\xi}  \; \langle L_1(z_2)
L_1(z_4) \rangle_{\mu\xi} + \langle L_1(z_1) L_1(z_4)
\rangle_{\mu\xi} \; \langle L_1(z_2) L_1(z_3) \rangle_{\mu\xi}.
\label{loop1}\eea
The contractions $\langle L_1(z_1) L_1(z_2) \rangle_{\mu\xi}  \;
\langle L_1(z_3) L_1(z_4) \rangle_{\mu\xi}$ do not appear above
since they do not contribute to one-loop corrections to the four-point
function. Collecting the above results and using Fourier transforms,
we have the first contribution to Eq. (\ref{loop}):
\begin{eqnarray}
\langle G(x_1,x_2)G(x_3,x_4) \rangle^{(1-Ia)}_{\mu\xi} &=& \int
\frac{dk_1}{(2\pi)^{d+1}} \cdots \int
\frac{dk_6}{(2\pi)^{d+1}}\,e^{-i( k_1 x_1 - k_2 x_2 + k_3 x_3 - k_4
x_4)} G^{(0)}(k_1) G^{(0)}(k_2) G^{(0)}(k_3)
G^{(0)}(k_4)\nn\\
&\times& \, G^{(0)}(k_5) G^{(0)}(k_6)\left(\sigma^2_\mu \omega^2_2
\omega^2_4 + \sigma^2_\xi m^4_0\right)\left(\sigma^2_\mu \omega^2_5
\omega^2_6 + \sigma^2_\xi m^4_0\right)\nn\\
&\times& \,
 \int dz_1 \cdots \int dz_4 \, e^{i(k_1 - k_5)z_1 + i(k_5 - k_2)z_2 +
 i(k_3 - k_6)z_3 + i(k_6 - k_4)z_4}\delta(\z_1 - \z_3)\delta(\z_2 - \z_4)
\nn\\ &=&\int\frac{dk_1}{(2\pi)^{d+1}} \cdots \int
\frac{dk_6}{(2\pi)^{d+1}}e^{-i( k_1 x_1 + k_2 x_2 + k_3 x_3 + k_4
x_4)} G^{(0)}(k_1) G^{(0)}(k_2) G^{(0)}(k_3)
G^{(0)}(k_4)\nn\\
&\times& \, G^{(0)}(k_5) G^{(0)}(k_6)\left(\sigma^2_\mu \omega^2_2
\omega^2_4 + \sigma^2_\xi m^4_0\right)\left(\sigma^2_\mu \omega^2_5
\omega^2_6 + \sigma^2_\xi m^4_0\right)(2\pi)^{d+1} \delta(k_1 + k_2
+ k_3 + k_4)\nn\\
&\times& \,(2\pi)^{d+1} \delta(k_2 + k_4 + k_5 + k_6)
(2\pi)\delta(\omega_3 + \omega_4)(2\pi)\delta(\omega_2 + \omega_5).
\label{Ia}
\end{eqnarray}
To reach the last result we used the identities:
\begin{equation}
\delta(-\k_1 - \k_3 + \k_5 + \k_6)\delta(\k_2 + \k_4 - \k_5 - \k_6)
= \delta(-\k_1 + \k_2 - \k_3 + \k_4)\delta(\k_2 + \k_4 - \k_5 -
\k_6),
\end{equation}
\begin{eqnarray}
&&\delta(\omega_1 - \omega_5)\delta(\omega_5 - \omega_2)
\delta(\omega_3 - \omega_6)\delta(\omega_6 - \omega_4) =
\delta(\omega_1 - \omega_2)\delta(\omega_3 - \omega_4)
\delta(\omega_5 - \omega_2)\delta(\omega_6 - \omega_4) \nn \\ &=&
\delta(\omega_1 - \omega_2 + \omega_3 - \omega_4)\delta(\omega_3 -
\omega_4)\delta(-\omega_2 - \omega_4 + \omega_5 +
\omega_6)\delta(\omega_5 - \omega_2),
\end{eqnarray}
and, finally
\begin{eqnarray}
&&\delta(\omega_1 - \omega_2 + \omega_3 - \omega_4)\delta(\omega_3 -
\omega_4)\delta(-\omega_2 - \omega_4 + \omega_5 +
\omega_6)\delta(\omega_5 - \omega_2)\delta(-\k_1 + \k_2 - \k_3 +
\k_4)\delta(\k_2 + \k_4 - \k_5 - \k_6) \nn \\ &=& \delta(-k_1 + k_2
- k_3 + k_4)\delta(k_2 + k_4 - k_5 - k_6)\delta(\omega_3 -
\omega_4)\delta(\omega_5 - \omega_2)
\end{eqnarray}
Then, we have made the changes $k_2 \rightarrow - k_2$ and $k_4
\rightarrow - k_4$.

As for the second contribution to equation (\ref{loop}), we have:
\begin{eqnarray}
\langle G(x_1,x_2)G(x_3,x_4) \rangle^{(1-Ib)}_{\mu\xi} &=& \int
\frac{dk_1}{(2\pi)^{d+1}} \cdots \int
\frac{dk_6}{(2\pi)^{d+1}}\,e^{-i( k_1 x_1 - k_2 x_2 + k_3 x_3 - k_4
x_4)} G^{(0)}(k_1) G^{(0)}(k_2) G^{(0)}(k_3)
G^{(0)}(k_4)\nn\\
&\times& \, G^{(0)}(k_5) G^{(0)}(k_6)\left(\sigma^2_\mu \omega^2_2
\omega^2_6 + \sigma^2_\xi m^4_0\right)\left(\sigma^2_\mu \omega^2_4
\omega^2_5 + \sigma^2_\xi m^4_0\right)\nn\\
&\times& \,
 \int dz_1 \cdots \int dz_4 \, e^{i(k_1 - k_5)z_1 + i(k_5 - k_2)z_2 +
 i(k_3 - k_6)z_3 + i(k_6 - k_4)z_4}\delta(\z_1 - \z_3)\delta(\z_2 - \z_4)
\nn\\ &=&\int\frac{dk_1}{(2\pi)^{d+1}} \cdots \int
\frac{dk_6}{(2\pi)^{d+1}}e^{-i( k_1 x_1 + k_2 x_2 + k_3 x_3 + k_4
x_4)} G^{(0)}(k_1) G^{(0)}(k_2) G^{(0)}(k_3)
G^{(0)}(k_4)\nn\\
&\times& \, G^{(0)}(k_5) G^{(0)}(k_6)\left(\sigma^2_\mu \omega^2_2
\omega^2_6 + \sigma^2_\xi m^4_0\right)\left(\sigma^2_\mu \omega^2_4
\omega^2_5 + \sigma^2_\xi m^4_0\right)(2\pi)^{d+1} \delta(k_1 + k_2
+ k_3 + k_4)\nn\\
&\times& \,(2\pi)^{d+1} \delta(k_2 + k_4 + k_5 + k_6)
(2\pi)\delta(\omega_3 + \omega_4)(2\pi)\delta(\omega_2 + \omega_5).
\label{Ib}
\end{eqnarray}
The last step in the above equation was achieved through the
identities:
\begin{equation}
\delta(-\k_1 + \k_4 + \k_5 - \k_6)\delta(\k_2 - \k_3 - \k_5 + \k_6)
= \delta(-\k_1 + \k_2 - \k_3 + \k_4)\delta(\k_2 - \k_3 - \k_5
+\k_6),
\end{equation}
\begin{eqnarray}
&&\delta(\omega_1 - \omega_5)\delta(\omega_5 - \omega_2)
\delta(\omega_3 - \omega_6)\delta(\omega_6 - \omega_4) =
\delta(\omega_1 - \omega_2)\delta(\omega_3 - \omega_4)
\delta(\omega_5 - \omega_2)\delta(\omega_3 - \omega_6) \nn \\ &=&
\delta(\omega_1 - \omega_2 + \omega_3 - \omega_4)\delta(\omega_3 -
\omega_4)\delta(-\omega_2 + \omega_3 + \omega_5 -
\omega_6)\delta(\omega_5 - \omega_2),
\end{eqnarray}
and, finally
\begin{eqnarray}
&&\delta(\omega_1 - \omega_2 + \omega_3 - \omega_4)\delta(\omega_3 -
\omega_4)\delta(-\omega_2 + \omega_3 + \omega_5 -
\omega_6)\delta(\omega_5 - \omega_2)\delta(-\k_1 + \k_2 - \k_3 +
\k_4)\delta(\k_2 - \k_3 - \k_5 +\k_6) \nn \\ &=& \delta(-k_1 + k_2 -
k_3 + k_4)\delta(k_2 - k_3 - k_5 + k_6)\delta(\omega_3 -
\omega_4)\delta(\omega_5 - \omega_2)
\end{eqnarray}
Similarly as before, we have also made the changes $k_2 \rightarrow
- k_2$, $k_4 \rightarrow - k_4$ and $k_6 \rightarrow - k_6$.

The second kind of contractions that also lead to one-loop corrections to the four-point function is shown in
Fig. \ref{gra_apend}.  For the first term in Eq. (\ref{g4}), we have:
\bea \langle G(x_1,x_2)G(x_3,x_4) \rangle^{(1-II)}_{\mu\xi} &=& \int
dz_1 \int dz_2 \, G^{(0)}(x_1,z_1) G^{(0)}(x_3,z_2) \langle\,{\cal
G}^{(1)}(z_1,x_2) {\cal G}^{(3)}(z_2,x_4)\,\rangle_{\mu\xi}.
\label{loop2}
\eea

From Eq. (\ref{genericterm}), one has:
\bea
\langle\,{\cal G}^{(1)}(z_1,x_2)
{\cal G}^{(3)}(z_2,x_4)\,\rangle_{\mu\xi} &=&
\int dz_{2} \int dz_4\, \langle L_1(z_1) L_1(z_2) L_1(z_3) L_1(z_4)
\rangle_{\mu\xi}
\nn\\
&\times&
\, G^{(0)}(z_1,x_2) G^{(0)}(z_2,z_3)  G^{(0)}(z_3,z_4) G^{(0)}(z_4,x_4).
\eea
Using again the fact that the noise correlations are Gaussian and since we are looking for one-loop corrections, we get:
\bea \langle L_1(z_1) L_1(z_2) L_1(z_3) L_1(z_4) \rangle_{\mu\xi}
&=& \langle L_1(z_1) L_1(z_3) \rangle_{\mu\xi}  \; \langle L_1(z_2)
L_1(z_4) \rangle_{\mu\xi}
\eea
The other contractions of the decomposition above do not appear
since they do not contribute to one-loop corrections to the four-point
function. Collecting the above results and using Fourier transforms,
we have:
\begin{eqnarray}
\langle G(x_1,x_2)G(x_3,x_4) \rangle^{(1-II)}_{\mu\xi} &=& \int
\frac{dk_1}{(2\pi)^{d+1}} \cdots \frac{dk_4}{(2\pi)^{d+1}}\int
\frac{dk}{(2\pi)^{d+1}}\frac{dk'}{(2\pi)^{d+1}}\,e^{-i( k_1 x_1 - k_2 x_2 + k_3 x_3 - k_4
x_4)} G^{(0)}(k_1) G^{(0)}(k_2)
\nn\\
&\times& \,  G^{(0)}(k_3)G^{(0)}(k_4)G^{(0)}(k) G^{(0)}(k')\left(\sigma^2_\mu \omega^2_2
\omega'^2 + \sigma^2_\xi m^4_0\right)\left(\sigma^2_\mu \omega^2
\omega^2_4 + \sigma^2_\xi m^4_0\right)\nn\\
&\times& \,
 \int dz_1 \cdots \int dz_4 \, e^{i(k_1 - k_2)z_1 + i(k_3 - k)z_2 +
 i(k - k')z_3 + i(k' - k_4)z_4}\delta(\z_1 - \z_3)\delta(\z_2 - \z_4)
\nn\\ &=&\int\frac{dk_1}{(2\pi)^{d+1}} \cdots \frac{dk_4}{(2\pi)^{d+1}} \int
\frac{dk}{(2\pi)^{d+1}}\frac{dk'}{(2\pi)^{d+1}}e^{-i( k_1 x_1 + k_2 x_2 + k_3 x_3 + k_4
x_4)} G^{(0)}(k_1) G^{(0)}(k_2)\nn\\
&\times& \,  G^{(0)}(k_3)
G^{(0)}(k_4)G^{(0)}(k) G^{(0)}(k')\left(\sigma^2_\mu \omega^2_2
\omega'^2 + \sigma^2_\xi m^4_0\right)\left(\sigma^2_\mu \omega^2
\omega^2_4 + \sigma^2_\xi m^4_0\right)(2\pi)^{d+1} \nn\\
&\times& \delta(k_1 + k_2
+ k_3 + k_4)\,(2\pi)^{d+1} \delta(k_1 + k_2 + k + k')
(2\pi)\delta(\omega_1 + \omega_2)(2\pi)\delta(\omega + \omega').
\label{II}
\end{eqnarray}
In the last equation we used similar identities of delta functions which were employed in calculation
of the expressions  (\ref{Ia}) and (\ref{Ib}). Just as
the tree-level case, all the one-loop contributions in equation (\ref{g4}) can be evaluated from the results in
equations (\ref{Ia}), (\ref{Ib}) and (\ref{II}) by adequate interchanges of variables.
Thus, the final result for the one-loop corrections for the four-point function,
$G^{(1)}_4(x_1,x_2,x_3,x_4)$, reads:
\bea G^{(1)}_4(x_1,x_2,x_3,x_4) &=& \int \frac{dk_1}{(2\pi)^{d+1}}
\cdots \int \frac{dk_4}{(2\pi)^{d+1}} (2\pi)^{d+1} \delta(k_1 + k_2
+ k_3 + k_4)\, \nn \\ &\times& e^{-i( k_1 x_1 + k_2 x_2 + k_3 x_3 +
k_4 x_4)}\, G^{(1)}_4(k_1,k_2,k_3,k_4), \eea
with
\bea G^{(1)}_4(k_1,k_2,k_3,k_4) = G^{(0)}(k_1) G^{(0)}(k_2)
G^{(0)}(k_3) G^{(0)}(k_4) \, \Gamma^{(1)}_4(k_1,k_2,k_3,k_4) , \eea
where
\bea \Gamma^{(1)}_4(k_1,k_2,k_3,k_4) &=&
  (2\pi)^{d+3}\int \frac{dk}{(2\pi)^{d+1}}\int \frac{dk'}{(2\pi)^{d+1}} G^{(0)}(k) G^{(0)}(k')
  \nn \\ &\times&\Bigl[ \delta(k_2 + k_4 + k + k')\delta(\omega_3 + \omega_4)
\delta(\omega + \omega_2) \,\left(\sigma^2_\mu \omega^2_2 \omega^2_4
+ \sigma^2_\xi m^4_0\right)\left(\sigma^2_\mu \omega^2 \omega'^2 +
\sigma^2_\xi m^4_0\right) \nn \\
&+& \delta(k_2 + k_3 + k + k')\delta(\omega_3 + \omega_4)
\delta(\omega + \omega_2)\left(\sigma^2_\mu \omega^2 \omega^2_4 +
\sigma^2_\xi m^4_0\right)\left(\sigma^2_\mu \omega'^2 \omega^2_2 +
\sigma^2_\xi m^4_0\right)
\nn \\
&+& \delta(k_3 + k_4 + k + k')\delta(\omega_2 + \omega_4)
\delta(\omega + \omega_3)\left(\sigma^2_\mu \omega^2_3 \omega^2_4 +
\sigma^2_\xi m^4_0\right)\left(\sigma^2_\mu \omega^2 \omega'^2 +
\sigma^2_\xi m^4_0\right)
\nn \\
&+& \delta(k_2 + k_3 + k + k')\delta(\omega_2 + \omega_4)
\delta(\omega + \omega_3)\left(\sigma^2_\mu \omega^2 \omega^2_4 +
\sigma^2_\xi m^4_0\right)\left(\sigma^2_\mu \omega'^2 \omega^2_3 +
\sigma^2_\xi m^4_0\right)
\nn \\
&+& \delta(k_3 + k_4 + k + k')\delta(\omega_2 + \omega_3)
\delta(\omega + \omega_4)\left(\sigma^2_\mu \omega^2_3 \omega^2_4 +
\sigma^2_\xi m^4_0\right)\left(\sigma^2_\mu \omega^2 \omega'^2 +
\sigma^2_\xi m^4_0\right)
\nn \\
&+& \delta(k_2 + k_4 + k + k')\delta(\omega_2 + \omega_3)
\delta(\omega + \omega_4)\left(\sigma^2_\mu \omega^2 \omega^2_3 +
\sigma^2_\xi m^4_0\right)\left(\sigma^2_\mu \omega'^2 \omega^2_4 +
\sigma^2_\xi m^4_0\right)
\nn \\
&+& \delta(k_1 + k_2 + k + k')\delta(\omega_1 + \omega_2)
\delta(\omega + \omega')\left(\sigma^2_\mu \omega^2_2 \omega'^2 +
\sigma^2_\xi m^4_0\right)\left(\sigma^2_\mu \omega^2 \omega^2_4 +
\sigma^2_\xi m^4_0\right)
\nn \\
&+& \delta(k_1 + k_3 + k + k')\delta(\omega_1 + \omega_3)
\delta(\omega + \omega')\left(\sigma^2_\mu \omega^2_3 \omega'^2 +
\sigma^2_\xi m^4_0\right)\left(\sigma^2_\mu \omega^2 \omega^2_4 +
\sigma^2_\xi m^4_0\right)
\nn \\
&+& \delta(k_1 + k_4 + k + k')\delta(\omega_1 + \omega_4)
\delta(\omega + \omega')\left(\sigma^2_\mu \omega^2_4 \omega'^2 +
\sigma^2_\xi m^4_0\right)\left(\sigma^2_\mu \omega^2 \omega^2_3 +
\sigma^2_\xi m^4_0\right)
\nn \\
&+& \delta(k_2 + k_3 + k + k')\delta(\omega_2 + \omega_3)
\delta(\omega + \omega')\left(\sigma^2_\mu \omega^2_3 \omega'^2 +
\sigma^2_\xi m^4_0\right)\left(\sigma^2_\mu \omega^2 \omega^2_4 +
\sigma^2_\xi m^4_0\right)
\nn \\
&+& \delta(k_2 + k_4 + k + k')\delta(\omega_2 + \omega_4)
\delta(\omega + \omega')\left(\sigma^2_\mu \omega^2_4 \omega'^2 +
\sigma^2_\xi m^4_0\right)\left(\sigma^2_\mu \omega^2 \omega^2_3 +
\sigma^2_\xi m^4_0\right)
\nn \\
&+& \delta(k_3 + k_4 + k + k')\delta(\omega_3 + \omega_4)
\delta(\omega + \omega')\left(\sigma^2_\mu \omega^2_4 \omega'^2 +
\sigma^2_\xi m^4_0\right)\left(\sigma^2_\mu \omega^2 \omega^2_2 +
\sigma^2_\xi m^4_0\right)
\Bigr]. \label{Gamma4-1}
\end{eqnarray}

\newpage

\end{widetext}


\vspace{1.0cm}



\begin{thebibliography}{99}
%
\bibitem{ford11} L. H. Ford, Phys. Rev. {\bf D51}, 1692 (1995).
%
\bibitem{fn} L. H. Ford and N. F. Svaiter, Phys. Rev. {\bf D56},
2226 (1997); L. H. Ford and N. F. Svaiter, Phys. Rev. {\bf D54},
2640 (1996); H. Yu and L. H. Ford,  Phys. Rev. {\bf D60}, 084023 (1999);
R. T. Thompson and L. H. Ford,  Phys. Rev. {\bf D78}, 024014 (2008);
R. T. Thompson and L. H. Ford, Class. Quant. Grav. {\bf 25}, 154006 (2008);
H. Yu, N. F. Svaiter and L. H. Ford, Phys. Rev. {\bf D80}, 124019 (2009).
%
\bibitem{hu1} B. L. Hu and K. Shiokawa, Phys. Rev. {\bf D57}, 3474
(1998).
%
\bibitem{krein1} G. Krein, G. Menezes and N. F. Svaiter,
Phys. Rev. Lett. {\bf 105}, 131301 (2010).
%
\bibitem{ishimaru} A. Ishimaru, {\em Wave Propagation
and Scattering in Random Media} (Academic, New York, 1978).
%
\bibitem{book3} H. Gzyl, {\em Diffusion and Waves} (Kluwer
Academic Publishers, Netherlands, 2002).
%
\bibitem{book2} V. I Klyatskin {\em Dynamics of Stochastic
Systems} (Elsevier B. V., Amsterdam, 2005).
%
\bibitem{book} J. Pierre Fouque, J. Garnier, G. Papanicolaou and
K. Solna, {\em Wave Propagation and Time Reversal in Randomly
Layered Media} (Spinger Science and Bussiness Media, New York, 2007).
%
\bibitem{birrell} N. D. Birrell and P. C. Davis, {\em Quantum Fields in
Curved Space} (Cambridge University Press, New York, 1982).
%
\bibitem{debashish} D. Chowdhury, {\em Spin Glasses and Other Frustrated
Systems} (World Scientific, Singapore, 1986).
%
\bibitem{glass} M. M\'ezard, G. Parisi, and M.A. Virasoro, {\em Spin
Glass Theory and Beyond} (World Scientific, Singapore, 1987)
%
\bibitem{parisi1} G. Parisi, Lett. Math. Phys. {\bf 88}, 255
(2009), S. W. Edwards and P. W. Anderson, J. Phys. {\bf F4},
965 (1975), D. Sherrington and S. Kirkpatrick,
Phys. Rev. Lett. {\bf 35}, 1972 (1975).
%
\bibitem{lub} T. C. Lubensky, Phys. Rev {\bf B11},
3573 (1975),
G. Grinstein and A. Luther, Phys. Rev {\bf B13},
1329 (1976),
J. A. Hertz, Phys. Rev. {\bf B18}, 4875
(1978), S. K. Ma and J. Rudnick, Phys. Rev. Lett. {\bf 40}, 589 (1978),
D. Sherrington, Phys. Rev. {\bf B22}, 5553
(1980),
D. Sherrington, J. Phys.  {\bf C14}, L371
(1981).
%
\bibitem{unruh1} W. G. Unruh, Phys. Rev. Lett. {\bf 46}, 1351
(1981).
%
\bibitem{mario} {\em Artificial Black Holes}, edited by M.Novello, M. Visser,
and G. Volovick (World Scientific, Singapore, 2002).
%
\bibitem{lectures} {\em Quantum Analogues: From Phase Transitions
to Black Holes and Cosmology}, edited by R. Sch\"utzhold and W.G. Unruh,
Springer Lecture Notes in Physics Vol. 718 (Springer, Berlin Heidelberg,
2007).
%
\bibitem{am7} L. J. Garay, J. R. Anglin, J. I. Cirac and P. Zoller,
Phys. Rev. Lett. {\bf 85}, 4643 (2000); L. J. Garay, J. R. Anglin, J. I. Cirac
and P. Zoller,  Phys. Rev. {\bf A63}, 023611 (2001);  C. Barcel\'o, S. Liberati,
and M. Visser, Class. Quant. Grav. {\bf 18}, 1137 (2001); P. O. Fedichev and
U. R. Fisher,  Phys. Rev. Lett. {\bf 91}, 240407 (2003); V. A. De Lorenci, and
R. Klippert, Braz. J. Phys. {\bf 34}, 1367 (2004); H. Nakano, Y. Kurita, K. Ogawa,
and C. Moon Yoo, Phys. Rev. {\bf D71}, 084006 (2005); R. Sch\"utzhold, Phys. Rev.
Lett. {\bf 97}, 190405 (2006); S. Wuster and C. M. Savage, Phys. Rev. {\bf A76},
013608 (2007).
%
%
\bibitem{matsu} T. Matsubara, Prog. Theor. Phys. {\bf 55}, 351 (1955);
R. Kubo, J. Phys. Soc. Jap. {\bf 12}, 570 (1957); P. C. Martin and
J. Schwinger, Phys. Rev. {\bf 115}, 1342 (1959).
%
\bibitem{ford} L. H. Ford and N. F. Svaiter, Phys. Rev. {\bf D51}, 6981 (1995).
%
\bibitem{sy} K. Symanzik, Nucl. Phys. {\bf B190}, 1 (1981); C. D. Fosco and
N. F. Svaiter, J. Math. Phys. {\bf 42}, 5185, (2001); R. B. Rodrigues and
N. F. Svaiter, Physica {\bf A328}, 466 (2003); R. B. Rodrigues and N. F. Svaiter,
Physica {\bf A342}, 529 (2004); M. I. Caicedo and N. F. Svaiter, J. Math.
Phys. {\bf 45}, 179 (2004); N. F. Svaiter, J. Math. Phys. {\bf 45}, 4524 (2004);
M. Aparicio Alcalde, G. F. Hidalgo, and N. F. Svaiter, J. Math.
Phys. {\bf 47}, 052303 (2006).
%
\bibitem{arias} E. Arias, N. F. Svaiter, and G. Menezes, Phys. Rev. {\bf D82},
045001 (2010).
%
%
%
%
%
\bibitem{cluster} J.-P. Hansen and I. R. MacDonald,
{\em Theory of Simple Liquids} (Academic Press, London, 2008).
%
\bibitem{anderson} P. W. Anderson, Phys. Rev. {\bf 109}, 1492 (1958).
%
\bibitem{new} G. Krein, G. Menezes, T. C. de Aguiar, E. Arias and N. F. Svaiter
{\em Bose-Einstein Condensation in Inhomogeneous Random Media}, in preparation.
%
\bibitem{Fagnocchi:2010sn}
  S.~Fagnocchi, S.~Finazzi, S.~Liberati, M.~Kormos, and A.~Trombettoni,
  New J.\ Phys.\  {\bf 12}, 095012 (2010)
  [arXiv:1001.1044 [gr-qc]].
%
\end{thebibliography}
\end{document}